\documentclass[a4paper,12pt]{article}
\usepackage{graphicx,fullpage,cite}
\usepackage{amsfonts,amsmath,amssymb}
\usepackage{appendix}

\numberwithin{equation}{section}

\title{Generalized entropy production fluctuation theorems for quantum systems}
\author{Shubhashis Rana\footnote{email: shubho@iopb.res.in}, ~Sourabh Lahiri\footnote{email: lahiri@iopb.res.in} ~and A. M. Jayannavar\footnote{email: jayan@iopb.res.in}}
\date{}

\allowdisplaybreaks
\begin{document}

\newcommand{\nwc}{\newcommand}
\nwc{\la}{\langle}
\nwc{\ra}{\rangle}
\nwc{\lw}{\linewidth}
\nwc{\nn}{\nonumber}
\nwc{\Ra}{\Rightarrow}
\nwc{\dg}{\dagger}
\nwc{\td}{\tilde}
\nwc{\mI}{\mathcal{I}}

\nwc{\Tr}[1]{\underset{#1}{\mbox{\large Tr}}~}
\nwc{\pd}[2]{\frac{\partial #1}{\partial #2}}
\nwc{\ppd}[2]{\frac{\partial^2 #1}{\partial #2^2}}

\nwc{\zprl}[3]{Phys. Rev. Lett. ~{\bf #1},~#2~(#3)}
\nwc{\zpre}[3]{Phys. Rev. E ~{\bf #1},~#2~(#3)}
\nwc{\zpra}[3]{Phys. Rev. A ~{\bf #1},~#2~(#3)}
\nwc{\zprb}[3]{Phys. Rev. B ~{\bf #1},~#2~(#3)}
\nwc{\zjsm}[3]{J. Stat. Mech. ~{\bf #1},~#2~(#3)}
\nwc{\zepjb}[3]{Eur. Phys. J. B ~{\bf #1},~#2~(#3)}
\nwc{\zrmp}[3]{Rev. Mod. Phys. ~{\bf #1},~#2~(#3)}
\nwc{\zepl}[3]{Europhys. Lett. ~{\bf #1},~#2~(#3)}
\nwc{\zjsp}[3]{J. Stat. Phys. ~{\bf #1},~#2~(#3)}
\nwc{\zptps}[3]{Prog. Theor. Phys. Suppl. ~{\bf #1},~#2~(#3)}
\nwc{\zpt}[3]{Physics Today ~{\bf #1},~#2~(#3)}
\nwc{\zap}[3]{Adv. Phys. ~{\bf #1},~#2~(#3)}
\nwc{\zjpcm}[3]{J. Phys. Condens. Matter ~{\bf #1},~#2~(#3)}
\nwc{\zjpa}[3]{J. Phys. A ~{\bf #1},~#2~(#3)}
\nwc{\zpjp}[3]{Pram. J. Phys. ~{\bf #1},~#2~(#3)}
\nwc{\zpa}[3]{Physica A ~{\bf #1},~#2~(#3)}

\maketitle{}

\vspace{-0.7cm}
\begin{center}
Institute of Physics, Sachivalaya Marg, Bhubaneswar - 751005, India
\end{center}

\begin{abstract}
Based on trajectory dependent path probability formalism in state space,
 we derive generalized entropy production fluctuation relations for a quantum system in the 
presence of measurement and feedback. We have obtained these results for three different cases: 
(i) the system is evolving in isolation from its surroundings; (ii) the system being weakly coupled to a heat bath; and (iii) system in contact with reservoir using quantum Crooks fluctuation theorem. In case (iii), we build on the treatment carried out in [H. T. Quan and H. Dong, arxiv/cond-mat: 0812.4955], where a quantum trajectory has been defined as a sequence of alternating work and heat steps. The obtained entropy production fluctuation theorems retain the same form as in the classical case. The inequality of second law of thermodynamics gets modified in the presence of information. These fluctuation theorems are robust against intermediate measurements of any observable performed with respect to von Neumann projective measurements as well as weak or positive operator valued measurements.
\end{abstract}

\hspace{1cm}\small PACS: 05.40.Ca, 05.70.Ln, 03.65.Ta
\normalsize

\section{Introduction}
Nonequlibrium processes are common in nature, but a general framework to understand them is lacking as compared to equilibrium systems. However, recent development in the field of nonequilibrium statistical mechanics, had led for the discovery of {\it fluctuation theorems} (FT) \cite{eva93,eva94,jar97,cro98,cro99,sei05,sei08}, which are exact equalities that are valid even when the system of interest is driven far away from equilibrium. For such a nonequlibrium system, the statistical distribution of thermodynamic quantities such as work, heat, entropy etc. exhibit universal relations. These thermodynamic quantities have now been generalized to a single trajectory of system evolving in phase space. They are random variables depending on the phase space trajectory (stochastic thermodynamics \cite{sek98}). The physical origin of FTs rely on the time reversal symmetry of the dynamics \cite{cro98,cro99} and they are expected to have important applications in nanoscience and biophysics. The second law of thermodynamics emerges in the form of inequalities from these theorems \cite{jar97,sei05}. It can be shown that second law is valid on average. Here averaging is done over different trajectories, thus not ruling out the possibility of transient violations of second law for individual realisation \cite{sah11}. These theorems have helped us in understanding how thermodynamic irreversibility arises from the underlying time reversible dynamics \cite{kaw07}.

One of the FT was initially put forward by Jarzynski \cite{jar97} in the form of the nonequilibrium work theorem, by means of which one can extract information about equilibrium changes in free energy $\Delta F$ by measuring the nonequilibrium work  $W$ performed on a system by the external drive. The system is initially prepared in equilibrium, and then driven away from equilibrium using some predetermined protocol $\lambda(t)$ which runs from $t=0$ to $t=\tau$. The Jarzynski Equality is given by
\begin{equation}
 \la e^{-\beta W}\ra = e^{-\beta \Delta F}.
\end{equation}
The work $W$, depends on trajectory of the system, whose initial state is sampled from equilibrium distribution. The angular brackets denote averaging over an ensemble of such trajectories and the free energy differences $\Delta F=F(\lambda(\tau))-F(\lambda(0)$. A stronger fluctuation theorem was provided by Crooks \cite{cro98,cro99} in the form
\begin{equation}
 \frac{P_f(W)}{P_r(-W)} = e^{\beta(W-\Delta F)},
\end{equation}
$P_f(W)$ and $P_r(W)$ being the work probability densities generated under the forward protocol $\lambda(t)$ and the reverse protocol $\lambda(\tau-t)$, respectively.

A more general FT was put forward by Seifert \cite{sei08} which contains the Jarzynski and the Crooks theorems as special cases. A system which is in contact with a heat bath, is initially prepared in some arbitrary distribution $p_0(x_0)$ of phase space points and is perturbed by varying an external parameter $\lambda(t)$  up to time $t=\tau$. In the reverse process, the system evolves from some other initial distribution $p_1(x_\tau)$ under the time-reversed protocol $\lambda(\tau-t)$. The Seifert's fluctuation theorem  states that, the probability of a phase space trajectory along the forward process, $P[x(t)]$ is related to that along the reverse process, $\tilde P[\tilde x(t)]$, as
\begin{equation}
 \frac{P[x(t)]}{\tilde P[\tilde x(t)]}=\frac{P[x(t)|x_0]p_0(x_0)}{\tilde P[\tilde x(t)|x_\tau]p_1(x_\tau)} = \frac{p_0(x_0)}{p_1(x_\tau)} \exp[{\Delta S_B}],
\end{equation}
where $\Delta S_B$ is the change in the entropy of the bath ($\Delta S_B =\frac{Q}{T}$, where Q is heat absorbed by the bath). $P[x(t)|x_0]$ is a short notation for functional of a path starting at $x_0$ and $x(t)$ being the phase space trajectory ending at $x_\tau$. If, in particular, the distribution  $p_1(x_\tau)$ in the final distribution at time $\tau$ as dictated by dynamics, then the above relation gives the integral fluctuation theorem (IFT) for total entropy production \cite{sei08}:
\begin{equation}
 \la e^{-\Delta s_{tot}}\ra = 1.
\label{IFT}
\end{equation}
where 
\begin{equation}
 \Delta S_{tot}=\Delta S + \Delta S_B= \ln\frac{p_0(x_0)}{p_1(x_\tau)} + \frac{Q}{T}.
\label{totent}
\end{equation}
 Here $\ln\frac{p_0(x_0)}{p_1(x_\tau)}$ is the change of system entropy along a given trajectory. For details we refer to Seifert's article \cite{sei08}. If the system is in steady state, one can also obtain detailed entropy production fluctuation theorem (DFT), namely 
\begin{equation}
 \frac{p(\Delta S_{tot})}{p(-\Delta S_{tot})} = e^{\Delta S_{tot}}.
\end{equation}
The IFT follows directly from the DFT.
Using  Jensen's inequality in Eq.(\ref{IFT}) we get
\begin{equation}
 \la\Delta S_{tot}\ra \ge 0.
\end{equation}
This is a statement of second law of thermodynamics, expressed in the form of inequality for the average change in total entropy.

If the systems are driven by the feedback controlled protocols, which in turn depend on the measurement outcomes of the state of the system at intermediate times (information gain), then IFT  gets modified a form \cite{lah12}
\begin{equation}
  \la e^{-\Delta S_{tot}-I}\ra = 1,
\label{ModIFT}
\end{equation}
where $I$ is the mutual information which quantifies the change in uncertainty of state of the system upon making measurements. Application of Jensen's inequality generalizes second law for total entropy production:
\begin{equation}
 \la \Delta S_{tot}\ra \ge -\la I\ra.
\label{ModSL}
\end{equation}
The average mutual information $\la I\ra$, is always non negative\cite{cov}. Thus the average entropy change can be made negative by feedback control, and the lower bound is given by  $-\la I\ra$. There are few attempts to extend the IFT (Eq.(\ref{IFT})) to the quantum domain \cite{muk06,mon05,lut12}. In our present work we extend the IFT for  $\Delta S_{tot}$ to quantum systems, in presence of multiple measurements and feedback. We assume that measurement procedure involves errors that are classical in nature. We  show the robustness of FTs against intermediate measurements of any system observable (both von Neumann projective measurements or generalized positive operator valued measurements (POVM)).

 We obtain these theorems for three different cases: (i) the system evolves in isolation from its surroundings; (ii) it is weakly coupled to a heat bath; and (iii) evolution of system coupled to heat bath is modelled in terms of work steps and heat steps following closely the treatment given Ref \cite{qua08} and is described in section \ref{strong}. Our treatment is based on path probability in state space. The measurement is assumed to be von Neumann type, i.e, projective measurement which results in the collapse of system state to one of the eigenstates of the corresponding observable. Case (i) namely isolated quantum system is discussed in detail. DFT is obtained for  various different situations, i.e., (a) system evolving unitarily, (b) in the presence of measurement and feedback,  and finally (c) in the presence of intermediate measurements, of any observables of the system. The IFT follows from DFT. For cases (ii) and (iii) we have derived generalized IFT. In the appendix, we have given a proof of IFT in presence of weak measurements. In passing, we note that all the extended quantum FTs retain the same form as their classical counterparts.
\section{Isolated quantum system}
\label{isoquant}
\subsection{Unitary evolution}
\label{uevol}

In this section we consider an isolated quantum system given by Hamiltonian $H(\lambda(t))$, where $\lambda(t)$ is some external time dependent protocol. To clarify our notation and for completeness we rederive DFT for this system following the treatment of ref \cite{mon05}. Initially at time t=0 energy measurement is performed and system is found to be in  eigenstate $|i_0\ra$, with energy eigenvalue $E_{0}$. It then evolves unitarily from time $0$ to $\tau$ under the protocol $\lambda(t)$. The energy measurement at final time $\tau$ is performed  and system is found to be at state $|i_{\tau}\ra$ with energy eigenvalue $E_{\tau}$. If the initial probability density of the state $|i_0\ra$ is $p(i_0)$ then the joint probability of $|i_0\ra$ and $|i_{\tau}\ra$ (forward state trajectory) is given by
\begin{align}
 P_F(i_{\tau},i_{0})
&=p(i_{\tau}|i_{0})p(i_{0})\nn\\
&=|\langle i_{\tau}|U_{\lambda}(\tau,0)|i_{0}\rangle|^{2}p(i_{0}),
%\label{jprob2}
\end{align}
where $U_{\lambda}(t_2,t_1)$ denotes the unitary evolution operator for given  $\lambda(t)$ from time $t_1$ to time $t_2$. It is defined as
\begin{equation}
 U_{\lambda}(t_{2},t_{1})=T \exp\left( -\frac{i}{\hbar}\int^{t_{2}}_{t_{1}}H(\lambda(t))dt\right).
\end{equation}
Here, $T$ denotes time ordering. 

 The system entropy is defined as $S(t)=-\ln p(i_{t})$. As the system is isolated there is no generation of heat, i.e, Q=0, Using eq. (\ref{totent}) the total change in entropy production $\Delta S_{tot}$ during the evolution from time 0 to $\tau$ is equal to the change in system entropy alone.
\begin{equation}
 \Delta S_{tot}=-\ln\frac{p(i_{\tau})}{p(i_{0})},
\label{totent1}
\end{equation}
where $p(i_{\tau})$  is the final probability of state $|i_{\tau}\ra$ at time $\tau$. 
 The probability density $P_F(\Delta S_{tot})$ for the forward path is by definition
\begin{align}
 P_F(\Delta S_{tot})&=\sum_{i_{\tau},i_{0}}\delta\left( \Delta S_{tot}+\ln\frac{p(i_{\tau})}{p(i_{0})}\right)  P_F(i_{\tau},i_{0})\nn\\
&=\sum_{i_{\tau},i_{0}}\delta\left( \Delta S_{tot}+\ln\frac{p(i_{\tau})}{p(i_{0})}\right)p(i_{\tau}|i_{0})p(i_{0}).
\label{iso1}
\end{align}
We now introduce time reversal operator $\Theta$. The time reversed state of  $|i\ra$ is defined as $|\td{i}\ra=\Theta|i\ra$.
It can be readily shown that \cite{lah12}
\begin{equation}
 p(i_2|i_1)=|\langle i_{2}|U_{\lambda}(t_{2},t_1)|i_{1}\rangle|^{2}=|\langle \tilde{i}_{1}|U_{\lambda^{\dg}}(\tilde{t}_{1},\tilde{t}_{2})|\tilde{i}_{2}\rangle|^{2}=p(\td{i}_1|\td{i}_2).
\label{microrev}
\end{equation}
where $\td{t}=\tau-t$ and $\lambda^{\dg}(\td{t})=\lambda(\tau-t)$ is the time reversed protocol of $\lambda(t)$.  The evolution of the system from given time reversed state $\Theta|i_2\ra$ to the time-reversed state $\Theta |i_1\ra$, under the time reversed protocol $\lambda^\dagger(t)$, is given by the conditional probability  $p(\td{i}_1|\td{i}_2)$. We consider the initial distribution of reverse trajectory to be equal to the final distribution of forward trajectory
\begin{equation}
 p(\td{i}_{\tau})= p(i_{\tau}).
\label{revprob}
\end{equation}
The states $|i\ra$ and $|\td{i}\ra$ have one-to-one correspondence.  Multiplying and dividing by $p(i_\tau)$ in the summand in eq.(\ref{iso1}) and using (\ref{microrev}) and (\ref{revprob}), we get
\begin{align}
P_F(\Delta S_{tot})&=\sum_{{i}_{\tau},{i}_{0}}\delta\left( \Delta S_{tot}+\ln\frac{p(i_{\tau})}{p(i_{0})}\right)p(\td{i}_{0}|\td{i}_{\tau})p(\td{i}_{\tau})\frac{p(i_{0})}{p(i_{\tau})}\nn\\
&=\sum_{{i}_{\tau},{i}_{0}}\delta\left( \Delta S_{tot}+\ln\frac{p(i_{\tau})}{p(i_{0})}\right)p(\td{i}_{0}|\td{i}_{\tau})p(\td{i}_{\tau})e^{\Delta S_{tot}}\nn\\
&=e^{\Delta S_{tot}}\sum_{{i}_{\tau},{i}_{0}}\delta\left( \Delta S_{tot}+\ln\frac{p(\td{i}_{\tau})}{p(\td{i}_{0})}\right)p(\td{i}_{0}|\td{i}_{\tau}) p(\td{i}_{\tau})\nn\\
&=e^{\Delta S_{tot}}\sum_{{i}_{\tau},{i}_{0}}\delta\left( \Delta S_{tot}-\ln\frac{p(\td{i}_{0})}{p(\td{i}_{\tau})}\right)P_R(\td{i}_{\tau},\td{i}_{0})\nn\\
&= e^{\Delta S_{tot}} P_R(-\Delta S_{tot}).
\label{asd}
\end{align}
To arrive at this result we have used eq.(\ref{totent1}) in the second step and eq.(\ref{revprob}) in third step. $P_R(\td{i}_{\tau},\td{i}_{0})$ is the joint probability of the corresponding states in the reverse direction. If the $\Delta S_{tot}$ is the total entropy change for forward path then the total entropy change in the corresponding reverse path is $-\Delta S_{tot}$. It follows from the fact that $p(\td{i}_\tau)$ and $p(\td{i}_0)$ is the initial and final probability distribution of state in the time reversed process because of unitary evolution. The eq.(\ref{asd}) can be written in the form
\begin{equation}
 \dfrac{P_F(\Delta S_{tot})}{P_R(-\Delta S_{tot})}=e^{\Delta S_{tot}}.
\end{equation}
This is the detailed fluctuation theorem for change in total entropy, extended to the quantum regime. Simple cross multiplication followed by integration over $\Delta S_{tot}$  leads to the integral form of the above theorem: 
\begin{equation}
 \langle e^{-\Delta S_{tot}}\rangle =1.
\end{equation}

\subsection{Isolated quantum system with feedback}
\label{isofeed}

So far we have been dealing with a predetermined protocol, also known as open loop feedback. Often to increase the efficiency of a physical process ( eg. engines at nanoscale, molecular motor etc.), we need to perform intermediate measurements and change the protocol as per the outcomes of these measurements \cite{jac03,jac04,jac06,jac08,sag10,sag11,pon10,hor10}. Such a process is known as closed loop feedback.
Let the system  evolve under some external protocol $\lambda_0(t)$, from its initial energy eigenstate $|i_0\ra$ measured at time $t_0$. At time $t_1$, we perform a measurement of some arbitrary observable and system collapses to state $|i_1\ra$. 
We assume that measurement process leading to information gain involves classical errors. Here $y_1$ is the measured outcome with a probability $p(y_1|i_1)$, while system's actual state is $|i_1\ra$. Depending on the value of $y_1$ the protocol is changed to $\lambda_{y_1}(t)$. Under this new protocol the system evolves unitarily up to time $t_2$ where another measurement is performed and so on. This process terminates at time $\tau$ when the system collapses to its final energy eigenvalue $|i_{\tau}\ra$. We should note that initial and final measurements are energy measurements. The joint probability of the corresponding state trajectory for n number of intermediate measurements $y_1, y_2,\cdots y_n$ at times $t_1,t_2,\cdots t_n$ respectively is \cite{lah12}
\begin{align}
 P_F(i_{\tau},..,i_{1},i_{0},y_{n},..,y_{1})&=p(i_{\tau}|i_{n})\cdots p(y_2|i_2)p(i_2|i_1)p(y_1|i_1)p(i_{1}|i_{0})p(i_{0})\nn\\
&=|\langle i_{\tau}|U_{\lambda_{y_{n}}}(\tau,t_{1})|i_{n}\rangle|^{2}...p(y_{2}|i_{2}) |\langle i_{2}|U_{\lambda_{y_1}}(t_{2},t_{1})|i_{1}\rangle|^{2}p(y_{1}|i_{1}) \nn\\
&\hspace{4cm}\times|\langle i_{1}|U_{\lambda_0}(t_{1},0)|i_{0}\rangle|^{2}p(i_{0}).
\label{PFiso}
\end{align}
It may be noted that the joint probability of path is expressed using classical probability rules. This is because we perform projective measurement on the system which collapses to one of the eigenvalues of the measured observables \cite{lah12,ran12}. 
As a consequence, it wipes out the previous memory of evolution and the post-measurement evolution becomes uncorrelated to the pre-measurement evolution. 
Thus if one performs intermediate measurements along two paths, the interference effects between the two paths disappear and the quantum effects are suppressed. Hence in the presence of measurement, path probability in state space obeys classical probability rules, and is given by product of transition probability of  paths between consecutive measurements. However, it may be noted that quantum mechanics enters through the explicit calculation of transition probability between two consecutive states.

To generate the reverse trajectory of a path in state space given in eq.(\ref{PFiso}), we first choose one of the forward protocols with probability $p(y_n \cdots,y_2,y_1)$, and then blindly time reverse the protocol. We perform measurements at the appropriate times along reverse path to allow the state to collapse to the corresponding time-reversed eigenstates. We do not use these measurements to perform any feedback to respect causality \cite{hor10}. Then the expression for the joint probability of reverse trajectory is  given by
\begin{equation}
P_R(\td{i}_{\tau}\cdots ,\td{i}_{0},y_{n},..,y_{1}) = p(\td{i}_N|\td{i}_{\tau})\cdots p(\td{i}_0|\td{i}_1) p(i_{\tau})p(y_n\cdots ,y_1).
\label{PRisofed}
\end{equation}
The mutual information gain due to measurements between the measured values and the actual value is defined as \cite{lah12,hor10}
\begin{equation}
I=\ln \frac{p(y_n|i_n)...p(y_2|i_2)p(y_1|i_1)}{p(y_n \cdots,y_2,y_1)}.
\label{mutlinf}
\end{equation}
We now calculate the joint probability density $P_F(\Delta S_{tot},\mI)$ of the entropy production and $\mI$ along the forward path, which is 
\begin{align}
  P_F(\Delta S_{tot},\mathcal{I})&=\int dy_n\cdots dy_1\sum_{i_{\tau}\cdots,i_{1},i_{0}}\delta\left( \Delta S_{tot}+\ln\frac{p(i_{\tau})}{p(i_{0})}\right)\delta\left(\mathcal{I}-I(i_{n},..,i_{1},y_{n},..,y_{1})\right)\nn\\
&\hspace{5cm} \times P_F(i_{\tau},..,i_{1},i_{0},y_{n},..,y_{1})\nn\\
&=\int dy_n\cdots dy_1\sum_{i_{\tau}\cdots,i_{1},i_{0}}\delta\left( \Delta S_{tot}+\ln\frac{p(i_{\tau})}{p(i_{0})}\right)\delta\left(\mathcal{I}-I(i_n,..,i_{1},y_{n},..,y_{1})\right) \nn\\
&\hspace{5cm} \times p(i_{\tau}|i_{n})\cdots p(y_2|i_2)p(i_2|i_1)p(y_1|i_1)p(i_{1}|i_{0})p(i_{0})\nn\\ 
&=\int dy_n\cdots dy_1\sum_{i_{\tau}\cdots,i_{1},i_{0}}\delta\left( \Delta S_{tot}+\ln\frac{p(i_{\tau})}{p(i_{0})}\right)\delta\left(\mathcal{I}-I(i_n,..,i_{1},y_{n},..,y_{1})\right)\nn\\ 
& \hspace{5cm} \times p(\td{i}_N|\td{i}_{\tau})\cdots p(\td{i}_0|\td{i}_1) p(i_{\tau})p(y_n\cdots ,y_1)e^{\Delta S_{tot}+I}\nn\\ 
&=\int dy_n\cdots dy_1\sum_{i_{\tau}\cdots,i_{1},i_{0}}\delta\left( \Delta S_{tot}+\ln\frac{p(i_{\tau})}{p(i_{0})}\right)\delta\left(\mathcal{I}-I(i_n,..,i_{1},y_{n},..,y_{1})\right)\nn\\
&\hspace{5cm} \times P_R(\td{i}_{\tau}\cdots ,\td{i}_{0},y_{n},..,y_{1})
e^{\Delta S_{tot}+I}\nn\\ 
&=e^{\Delta S_{tot}+\mI}\int dy_n\cdots dy_1\sum_{i_{\tau}\cdots,i_{1},i_{0}}\delta\left( \Delta S_{tot}+\ln\frac{p(i_{\tau})}{p(i_{0})}\right)\delta\left(\mathcal{I}-I(i_n,..,i_{1},y_{n},..,y_{1})\right)\nn\\
&\hspace{5cm} \times P_R(\td{i}_{\tau}\cdots ,\td{i}_{0},y_{n},..,y_{1}) \nn\\
&= e^{\Delta S_{tot}+\mI} P_R(-\Delta S_{tot},\mathcal{I}).
 \end{align}
In deriving above result we have made use eq.(\ref{PFiso}), (\ref{PRisofed}), (\ref{mutlinf}). The path variable $I(i_n,..,i_{1},y_{n},..,y_{1})$ is given by eq. (\ref{mutlinf}), and $\mathcal{I}$ denotes its value. 
It is important to note that the probability density function $P_R(-\Delta S_{tot},\mI)$ gives the probability of reverse trajectories along which the entropy chage is $-\Delta S_{tot}$ and whose corresponding forward trajectory has the mutual information $\mathcal{I}$ between its measured outcomes and actual states.
Once again, the initial and final distributions of states along forward trajectory get interchanged in the reverse trajectory because of unitary evolution between measurements. Along the reverse trajectory  the change in total entropy is $-\Delta S_{tot}$. Thus we obtain the DFT
\begin{equation}
 \dfrac{P_F(\Delta S_{tot},\mI)}{P_R(-\Delta S_{tot},\mI)}=e^{\Delta S_{tot}+\mI}.
\end{equation}
From the above equation the extended version of IFT and second law, eqs. (\ref{ModIFT}) and (\ref{ModSL}) can be readily obtained as discussed in earlier subsection.
\subsection{Isolated system under multiple measurements}
In this subsection we restrict ourselves on the influence of intermediate measurements of arbitrary observables on the statistics of $\Delta S_{tot}$. To this end we do not involve any feedback. Following closely the discussions in section (\ref{isofeed}), 
 of path probability in state space is given by 
\begin{align}
 P(i_{\tau},..,i_{1},i_{0})&=p(i_{\tau}|i_{n})\cdots p(i_2|i_1)p(i_{1}|i_{0})p(i_{0})\nn\\
&=|\langle i_{\tau}|U_{\lambda_{y_{n}}}(\tau,t_{1})|i_{n}\rangle|^{2}... |\langle i_{2}|U_{\lambda_{y_1}}(t_{2},t_{1})|i_{1}\rangle|^{2}
|\langle i_{1}|U_{\lambda_0}(t_{1},0)|i_{0}\rangle|^{2}p(i_{0}).
\end{align}
 From preceding section we now calculate the probability density $P_F(\Delta S_{tot})$ of the total entropy change along forward path
\begin{align}
P_F(\Delta S_{tot})&=\sum_{i_{\tau}\cdots,i_{1},i_{0}}\delta\left( \Delta S_{tot}+\ln\frac{p(i_{\tau})}{p(i_{0})}\right)P(i_{\tau},..,i_{1},i_{0})\nn\\
&=\sum_{i_{\tau}\cdots,i_{1},i_{0}}\delta\left( \Delta S_{tot}+\ln\frac{p(i_{\tau})}{p(i_{0})}\right) p(i_{\tau}|i_{n})\cdots p(i_2|i_1)p(i_{1}|i_{0})p(i_{0})\nn\\ 
&=\sum_{i_{\tau}\cdots,i_{1},i_{0}}\delta\left( \Delta S_{tot}+\ln\frac{p(i_{\tau})}{p(i_{0})}\right)p(\td{i}_N|\td{i}_{\tau})\cdots p(\td{i}_0|\td{i}_1) p(i_{\tau})e^{\Delta S_{tot}}\nn\\ 
&=\sum_{i_{\tau}\cdots,i_{1},i_{0}}\delta\left( \Delta S_{tot}+\ln\frac{p(i_{\tau})}{p(i_{0})}\right)P_R(\td{i}_{\tau}\cdots ,\td{i}_{0})
e^{\Delta S_{tot}}\nn\\ 
&=e^{\Delta S_{tot}}\sum_{i_{\tau}\cdots,i_{1},i_{0}}\delta\left( \Delta S_{tot}+\ln\frac{p(i_{\tau})}{p(i_{0})}\right)P_R(\td{i}_{\tau}\cdots ,\td{i}_{0}),
 \end{align}
where $P_R(\td{i}_{\tau}\cdots ,\td{i}_{0})$ is the probability of reverse path. The DFT for $\Delta S_{tot}$ follows from the above equation:
\begin{equation}
 \dfrac{P_F(\Delta S_{tot})}{P_R(-\Delta S_{tot})}=e^{\Delta S_{tot}}.
\label{asdf}
\end{equation}
We observe from eq.(\ref{asdf}) the robustness of this FT against intermediate measurements \cite{han10,han11}. It retains the same form as in the classical case. The path probability, however, gets modified in presence of measurements and statistics of $\Delta S_{tot}$ is strongly influenced by the intermediate measurements. In the next section we derive IFT in presence of feedback for a quantum system coupled weakly to a bath. In the appendix, we have shown that the IFT for $\Delta S_{tot}$ is also robust against weak or generalized intermediate measurements.
\section{Weakly coupled quantum system}

Consider a driven system which is weakly coupled to a bath. The total Hamiltonian will be 
\begin{equation}
 H(\lambda(t))=H_S (\lambda(t))+ H_B+ H_{SB}.
\end{equation}
The external time dependent drive $\lambda(t)$ only affects the system Hamiltonian $H_S (\lambda(t))$ while the bath Hamiltonian $H_B$ and interaction Hamiltonian $H_{SB}$ are time independent. As the system is weakly coupled it is assumed that $H_{SB}$ is negligibly small compared to $H_S (\lambda(t))$ and $ H_B$. 
Initially  the super-system (system+bath) is coupled to a large reservoir of inverse temperature $\beta$ \cite{han11,han09}. At time $t=0$ the large reservoir is decoupled from the super-system. Hence initially the super-system will be in a canonical distribution,
\begin{equation}
\rho(\lambda_0)=\dfrac{e^{-\beta H(\lambda(0))}}{Y(\lambda(0))},
\end{equation}
where $Y(\lambda(0))= \Tr{}e^{-\beta H(\lambda(0))}$. The system and the bath Hamiltonians  commute with each other, hence we can measure 
simultaneously the energy eigenstates for system as well as bath. At t=0, the measured energy eigenvalues of system and bath are denoted by $E_0^{S}$ and $E_0^B$, respectively. We perform N number of intermediate measurements of some arbitrary observable at time $t_1,t_2\cdots t_N$ between time 0 to ${\tau}$. Initially the protocol was $\lambda_0(t)$. At $t_1$ the measured output is $y_1$, while its actual state is $i_1$, with  probability $p(y_1|i_1)$. Now the protocol is changed to $\lambda_{y_1}(t)$ and system  evolves up to time $t_2$ . Again measurement is performed and protocol is changed according to the output at intermediate times and so on.  Finally at t=$\tau$ joint measurement is performed on system and bath Hamiltonians, and the measured eigenvalues are $E_\tau^S$ and $E_\tau^B$, respectively. The system-reservoir interaction energy can be neglected in the presence of weak coupling. Hence during the evolution process from time $t=0$ to $t=\tau$ for a single realization the change in the internal energy of the system is given by \cite{han09}
\begin{equation}
 \Delta U=E_{\tau}^S-E_0^{S}
\end{equation}
 and the heat dissipated to the bath is
\begin{equation}
 Q=E_{\tau}^B-E_0^{B}.
\end{equation}
If $i_0$ and $i_{\tau}$ denote initial and final system energy eigenstates, then system entropy change is 
\begin{equation}
\Delta S_{sys}=-\ln \frac{p(i_{\tau})}{p(i_0)}, 
\end{equation}
and the total entropy change is
\begin{equation}
 \Delta S_{tot}=\Delta S_{sys}+\Delta S_{B}=-\ln \frac{p(i_{\tau})}{p(i_0)}+\frac{Q}{T},
\end{equation}
where T is the temperature of the bath. The mutual information between the state trajectory $\{i_1,i_2,\cdots, i_N\}$ and the measurement trajectory $\{y_1,y_2,\cdots y_N\}$ is  
\begin{equation}
 I \equiv \ln\left[\frac{p(y_1|i_1)\cdots p(y_N|i_N)}{P(y_1,\cdots, y_N)}\right].
\end{equation}
Denoting initial and final states of the bath by $\alpha_0$ and $\alpha_{\tau}$, it can written from microscopic reversibility \cite{sag10,hor10}
\begin{equation}
 p(i_{\tau},\alpha_{\tau}|i_0,\alpha_0)=p(\td{i}_0,\td{\alpha_0}|\td{i}_{\tau},\td{\alpha}_{\tau}).
\label{microrev1}
\end{equation}
where $ p(i_{\tau},\alpha_{\tau}|i_0,\alpha_0)$ is the total transition probability for system and reservoir to evolve from state $|i_0,\alpha_0\ra$ to $|i_\tau,\alpha_\tau\ra$ under the full Hamiltonian. Here $|\tilde i,\tilde\alpha\ra \equiv \Theta|i,\alpha\ra$ is the time-reversed state of $|i,\alpha\ra$.
To generate the reverse trajectory, proper causal protocol has to be used which has been discussed in section \ref{isofeed}. Thus the forward and the reverse path probabilities of trajectories are respectively given by
\begin{equation}
 P_F(A\to B)= p(i_{\tau},\alpha_{\tau}|i_N,\alpha_N)\cdots p(y_1|i_1)p(i_{1},\alpha_{1}|i_0,\alpha_0)p(i_0,\alpha_0),
\label{P_F}
\end{equation}
\begin{equation}
 P_R(A\leftarrow B)=p(\td{i}_0,\td{\alpha_0}|\td{i}_1,\td{\alpha}_1)\cdots p(\td{i}_N,\td{\alpha_N}|\td{i}_{\tau},\td{\alpha}_{\tau}) p(\td{i}_{\tau},\td{\alpha}_{\tau})
p(y_1,y_2,\cdots y_N).
\label{P_R}
\end{equation}
The notations A and B denote initial and final values of the forward protocol, respectively. For reverse trajectory we have chosen the outcome of the forward trajectory with probability $p(y_1,y_2,\cdots y_N)$ and have blindly reversed the protocol, but performing measurements (without any feedback) at appropriate time instants. From (\ref{P_F}) and (\ref{P_R}) we get
\begin{align}
 \frac{ P_F(A\to B)}{P_R(A\leftarrow B)}&=\frac{ p(i_{\tau},\alpha_{\tau}|i_N,\alpha_N)\cdots p(y_1|i_1)p(i_{1},\alpha_{1}|i_0,\alpha_0)p(i_0,\alpha_0)}
{p(\td{i}_0,\td{\alpha_0}|\td{i}_1,\td{\alpha}_1)\cdots p(\td{i}_N,\td{\alpha_N}|\td{i}_{\tau},\td{\alpha}_{\tau}) p(\td{i}_{\tau},\td{\alpha}_{\tau})
p(y_1,y_2,\cdots y_N)}\nn\\
&=\frac{p(y_N|i_N)\cdots p(y_1|i_1)}{P(y_1,\cdots, y_N)}\frac{p(i_0,\alpha_0)}{p(\td{i}_{\tau},\td{\alpha}_{\tau})}\nn\\
&=e^{I}\hspace{0.2 cm} \frac{p(i_0)p(\alpha_0)}{p(\td{i}_{\tau})p(\td{\alpha}_{\tau})}\nn\\.
\label{ratio}
\end{align}
In arriving at (\ref{ratio}), we have used microreversibility (\ref{microrev1}) and we have assumed that the system and the bath are weakly coupled.
The joint probability of system and bath states is approximated as a product of individual state probabilities. 
Correction to this factorized initial state is at least of second order in system-bath interaction, and therefore they can be neglected in the limit of weak coupling.
The bath probability can be considered canonical with inverse temperature $\beta$. 
This leads to
\begin{align}
 \frac{ P_F(A\to B)}{P_R(A\leftarrow B)}&=e^{I}~e^{\Delta S_{sys}}\frac{e^{-\beta E_0^B}/Z_B}{e^{-\beta E_{\tau}^B}/Z_B}
=e^{I}~e^{\Delta S_{sys}}e^{Q/T}
=e^{\Delta S_{tot}+I}.
\label{IFT_weak}
\end{align}
A simple cross multiplication and integration over paths gives the extended IFT. It may be noted that in our framework  we can also obtain the DFT, provided the system either begins and ends in equilibrium or in the same nonequilibrium steady state \cite{cro99}. In the next section, we prove the same IFT for $\Delta S_{tot}$ by means of the method developed in \cite{qua08} by using the quantum mechanical generalization of the Crooks fluctuation theorem.

\section{IFT using quantum Crooks fluctuation theorem}
\label{strong}

We  consider the system to be coupled to a bath, but there is no assumption made in regard to the strength of the coupling. Each time step in the entire evolution is divided into two substeps. In first substep the protocol is changed while in second,  protocol is kept fixed and system relaxes by dissipation of heat.   The total evolution is divided into N steps. Each step starts at $t_{n}$ and ends at $t_{n+1}$, where $n=0,1,2,\cdots N-1$. We closely follow the treatment in \cite{qua08}.

For a quantum adiabatic process the protocol changes slowly and the system remains in same eigenstate in the work step. However, in the present case the  work step is almost instantaneous and the process is non adiabatic. As a consequence the eigenstates before and after work step may be different. The system starts to evolve under a predetermined protocol $\lambda_0$. For simplicity, let us consider the observable measured at intermediate times to be the Hamiltonian itself. It can be readily generalized to the case of other observables. We consider that the feedback is applied at the beginning of the each work step and we change the protocol subsequently according to the result obtained from the measurement, as discussed earlier.
The conditional probability $p(y_{n-1}|i_{n-1})$ denotes that the measured outcome is $y_{n-1}$ while the actual collapsed state is $|i_{n-1},\lambda_{n-1}\ra$, at the beginning of the $n^{th}$ work step. Within the ket notation, $i_{n-1}$ represents the state of the system and $\lambda_{n-1}$ is the value of the control parameter. After the measurement of $t_{n-1}$, the protocol is changed to $\lambda_n(y_{n-1})$ from $\lambda_{n-1}(y_{n-2})$. During the work step, the system evolves unitarily from $t_{n-1}$ to $t'_{n-1}$, where it is measured to be in state $|i'_{n-1},\lambda_n\ra$.
 The time taken in the work substep is considered to be too small for the system to relax. In the $n^{th}$ heat step, the system relaxes from state $|i'_{n-1},\lambda_n\ra$ to $|i_n,\lambda_n\ra$. Therefore, the path followed by the system in state space of the measured eigenstates from state $|i_0,\lambda_0=A\ra$ to $|i_\tau,\lambda_\tau=B\ra$ is represented as
$|i_0,\lambda_0\ra \to|i'_0,\lambda_1\ra \to|i_1,\lambda_1\ra \to |i'_1,\lambda_2\ra  \to \cdots \to |i_{N-1},\lambda_{N-1}\ra \to  |i'_{N-1},\lambda_{N}\ra \to |i_{N},\lambda_{N}\ra $.
Let $E(i_n,\lambda_n)$ be the energy eigenvalue of state $|i_n,\lambda_n\ra$.
By adding the contributions from all the work steps, the total work done on the system is given by
\begin{equation}
W=\sum_{n=0}^{N-1}\left[ E(i'_n,\lambda_{n+1})-E(i_n,\lambda_n)\right],  
\end{equation}
while heat dissipated into the bath is
\begin{equation}
Q=-\sum_{n=0}^{N-1}\left[ E(i_{n+1},\lambda_{n+1})-E(i'_{n},\lambda_{n+1})\right].  
\label{Q}
\end{equation}
The change  in internal energy of the system along the trajectory is
\begin{equation}
 \Delta E=Q+W= E(i_N,\lambda_N)-E(i_0,\lambda_0).
\end{equation}
As before, the mutual information is 
\begin{equation}
I=\ln \frac{p(y_n|i_n)...p(y_2|i_2)p(y_1|i_1)}{p(y_n \cdots,y_2,y_1)}.
\label{I}
\end{equation}
The forward and the reverse path probabilities are respectively given by
\begin{equation}
P_F(A\to B) =p(i_0,\lambda_0)\prod_{n=0}^{N-1}p(y_n|i_n)p_F(|i_n,\lambda_n\ra\to |i_n',\lambda_{n+1}\ra)~p_F(|i_n',\lambda_{n+1}\ra\to |i_{n+1},\lambda_{n+1}\ra).
\end{equation}
and
\begin{align}
 P_R(A\leftarrow B)&=p(i_N,\lambda_N)p(y_n \cdots,y_1) \prod_{n=0}^{N-1}p_R(|\tilde i_n,\lambda_n\ra\leftarrow |\tilde i_n',\lambda_{n+1}\ra) p_R(|\tilde i_n',\lambda_{n+1}\ra\leftarrow|\tilde i_{n+1},\lambda_{n+1}\ra).
\end{align}
As mentioned earlier, during the work step, the system can be regarded as an isolated quantum system and evolution is completely determined by the time-dependent Hamiltonian $H_S(\lambda(t))$. Thus the evolution is unitary.
Microscopic reversibility for work step gives \cite{qua08}
\begin{equation}
p_F(|i_n,\lambda_n\ra\to |i_n',\lambda_{n+1}\ra) = p_R(|\tilde i_n,\lambda_n\ra\leftarrow |\tilde i_n',\lambda_{n+1}\ra).
\label{micrev}
\end{equation}

The heat steps or relaxation steps are assumed to be microscopically reversible and obey
 the local detailed balance for all the fixed values of the external parameter $\lambda$.
 The detailed balance condition in relaxation substep implies
\begin{equation}
\frac{P_F(|i'_{n},\lambda_{n+1}\ra\to |i_{n+1},\lambda_{n+1}\ra)}{P_R(|\tilde i'_{n},\lambda_{n+1}\ra\leftarrow |\tilde i_{n+1},\lambda_{n+1}\ra)} = \exp[-\beta(E_{n+1},\lambda_{n+1}-E(i'_{n},\lambda_{n+1})].
\label{detbal}
\end{equation}
Using above two equations we get
\begin{align}
\frac{ P_F(A\to B)}{P_R(A\leftarrow B)} =\frac{p(y_n|i_n)...p(y_2|i_2)p(y_1|i_1)}{p(y_n \cdots,y_2,y_1)}\frac{p(i_0,\lambda_0)}{p(i_N,\lambda_N)}\prod^{N-1}_0 \exp[-\beta(E_{n+1},\lambda_{n+1}-E(i'_{n},\lambda_{n+1})].
\label{DFT}
\end{align}
The total entropy change along the trajectory, $\Delta S_{tot} = \Delta S+\Delta S_B$, which is a trajectory dependent random variable, 
where $\Delta S\equiv -\ln\frac{P(i_N,\lambda_N)}{P(i_0,\lambda_0)}$ is the change in system entropy, and $\Delta S_B\equiv Q/T$ is the entropy change of the bath, along a single trajectory. Using (\ref{Q}) and (\ref{I}), eq. (\ref{DFT}) simplifies to
\begin{align}
 \frac{ P_F(A\to B)}{P_R(A\leftarrow B)}
=e^{I}~e^{\Delta S_{sys}}e^{Q/T}
=e^{\Delta S_{tot}+I}
\end{align}
This immediately leads to the generalized integral fluctuation theorem for total entropy change,  in the presence of measurement and feedback.
 In the above derivation, we have taken measurements for feedback at the beginning of the work  steps for simplicity. These measurements can  be performed at any time in-between the work steps. The result will not be affected. It would only make the notations more complicated and would not provide any new physical insight. Feedback cannot be performed within the heat step which be definition requires protocol to be held constant.

 As in case (ii), the DFT for $\Delta S_{tot}$ can be obtained if the initial and final distributions are in the equilibrium or in the same nonequilibrium steady state \cite{cro99}.

\section{Conclusions}

Based on the path probability formulation in state space, we have derived generalized total entropy production fluctuation theorems for quantum systems in presence of measurement and feedback, for three different cases. They retain the same form as in classical case. The second law of thermodynamics gets modified in the presence of information and feedback (eq. (\ref{ModSL})). For isolated quantum system with feedback, we have derived the generalized DFT for the total entropy. For this case DFT retains the same form in presence of multiple measurements of any system observable, thus showing the robustness of these fluctuation theorems against measurements (von Neumann type or generalized measurements). For the case (ii) of a weakly coupled quantum system under feedback, we have derived the extended IFT for total entropy. In case (iii), we have derived the extended IFT for $\Delta S_{tot}$, using the quantum Crooks fluctuation theorem, where quantum trajectory is characterized by a sequence of alternating work and heat steps. IFT is valid for any initial arbitrary state of a system. DFT in cases (ii) and (iii) can be obtained only when the system either begins or ends in equilibrium or remains in the same nonequilibrium steady state. By using our approach, the generalized DFT can be proved, but we have not provided the details. The derivation of the robustness of the fluctuation theorems against intermediate measurements is given only for case (i), namely, for the isolated quantum system. Following the same treatment, the robustness of fluctuation theorems can be readily demonstrated for cases (ii) and (iii) as well.

In conclusion, we have generalized total entropy production fluctuation theorem in presence of feedback to the quantum domain using three different approaches.

 \vspace{1cm}
{\large \bf Acknowledgements}
\normalsize

\vspace{0.5cm}
One of us (AMJ) thanks DST, India for financial support.

%\newpage
\appendix

\begin{center}{\bf \Large Appendix}\end{center}
\section{Isolated system under weak measurements}
In this appendix we derive IFT for $\Delta S_{tot}$ under weak measurement (POVM), as opposed to projective von Neumann type measurements considered in sec.\ref{uevol}. We follow the mathametical treatment given in \cite{cam11}. For simplicity we consider only one weak measurement is performed at intermediate time. The genaralization to multiple weak measurements is straight forward.  
Consider an isolated quantum system is controlled externally through time dependent protocol $\lambda(t)$.  Initially at $t=0$ energy measurement is performed  and the system is found to be in state $|n,0\ra$ with probability  density $p_n$. The density matrix becomes 
\begin{equation}
 \rho_n(0^+)=\frac{\Pi_n^0 \rho_0 \Pi_n^0}{p_n}
\label{t0}
\end{equation}
where $\rho_0 $ is the density matrix of the system before measurement and $ \Pi_n^0$ denotes von Neumann projective measurement operator and $p_n=\Tr{}\Pi_n^0 \rho_0  $. The system then evolves unitarily up to time $t_1$ and a weak measurement is performed and we get the density matrix  
\begin{equation}
 \rho_n(t_1^+)=\sum_r M_r U_{\lambda}(t_1,0) \rho_n(0^+)U_{\lambda}^{\dg}(t_1,0) M_r^{\dg}.
\label{t1}
\end{equation}
$M_r$ is the weak measurement operator with property $\sum_r M_r M_r^{\dg}=1$.
The system undergoes further unitarily evolution and finally the projective measurement is performed and  the system is found to be in state $|m,\tau\ra$ at time $\tau$. The conditional probability $P_{\lambda}(m,\tau|n,0)$ for system initially in state $|n,0\ra$ and finally in state $|m,\tau\ra$ is given by
\begin{equation}
 P_{\lambda}(m,\tau|n,0)=\Tr{}\Pi_m^f U_{\lambda}(\tau,t_1) \rho_n(t_1^+)  U_{\lambda}^{\dg}(\tau,t_1).
\label{tau}
\end{equation}
Thus the probability of the change in total entropy, 
\begin{equation}
 \Delta S_{tot}=-\ln p_m+\ln p_n,
\end{equation}
for a given pre-determined protocol $\lambda(t)$ is
\begin{equation}
 P_{\lambda}(\Delta S_{tot})=\sum_{m,n}\delta(\Delta S_{tot}+\ln p_m-\ln p_n)P_{\lambda}(m,\tau|n,0)p_n.
\label{app1}
\end{equation}
where $p_m$ is the probability of the system to stay at the end of protocol at final time $\tau$.
The Fourier Transform of this probability is
\begin{align}
 G_{\lambda}(u)=\int d \Delta S_{tot} P_{\lambda}(\Delta S_{tot})e^{i u \Delta S_{tot}}.
\label{app3}
\end{align}
Substituting the expression for $P_{\lambda}(\Delta S_{tot})$ from eq.(\ref{app1}) and using eqs.(\ref{tau}),(\ref{t1}),(\ref{t0}) we get
\begin{align}
 G_{\lambda}(u)&=\sum_{m,n,r} e^{i u (-\ln p_m+\ln p_n)} \Tr{}\Pi_m^f U_{\lambda}(\tau,t_1) 
M_r U_{\lambda}(t_1,0) \Pi_n^0 \rho_0 \Pi_n^0 U_{\lambda}^{\dg}(t_1,0) M_r^{\dg}  U_{\lambda}^{\dg}(\tau,t_1)\nn\\
&=\sum_{m,n,r}  \Tr{}\Pi_m^f e^{-i u \ln \rho_f }U_{\lambda}(\tau,t_1)
M_r U_{\lambda}(t_1,0) \Pi_n^0 e^{i u \ln \rho_0 }\rho_0  U_{\lambda}^{\dg}(t_1,0) M_r^{\dg}  U_{\lambda}^{\dg}(\tau,t_1)\nn\\
&=\sum_{r}  \Tr{} e^{-i u \ln \rho_f }U_{\lambda}(\tau,t_1)
M_r U_{\lambda}(t_1,0)  e^{i u \ln \rho_0 }\rho_0  U_{\lambda}^{\dg}(t_1,0) M_r^{\dg}  U_{\lambda}^{\dg}(\tau,t_1)\nn\\
&=\sum_{r}  \Tr{} U_{\lambda}^{\dg}(t_1,0) M_r^{\dg}  U_{\lambda}^{\dg}(\tau,t_1)e^{-i u \ln \rho_f }U_{\lambda}(\tau,t_1)
M_r U_{\lambda}(t_1,0)  e^{i u \ln \rho_0 }\rho_0  \nn\\
\end{align}
$\rho_{f}$ is the final density matrix which is diagonal in the energy basis. In the second step we have used the completeness relation $\sum_{m} \Pi_m^f=1$ and $\sum_{n} \Pi_{n}^0=1$ for the projective operator. Substituting  $u=i$, for the Fourier Transform variable, we get from eq.(\ref{app3}) $G_{\lambda}(i)=\la e^{-\Delta S_{tot}}\ra$ and hence
\begin{align}
\la e^{-\Delta S_{tot}}\ra &= \sum_{r}  \Tr{} U_{\lambda}^{\dg}(t_1,0) M_r^{\dg}  U_{\lambda}^{\dg}(\tau,t_1)e^{ \ln \rho_f }U_{\lambda}(\tau,t_1)
M_r U_{\lambda}(t_1,0)  e^{- \ln \rho_0 }\rho_0  \nn\\
&= \sum_{r}  \Tr{} U_{\lambda}^{\dg}(t_1,0) M_r^{\dg}  U_{\lambda}^{\dg}(\tau,t_1)e^{ \ln \rho_f }U_{\lambda}(\tau,t_1)
M_r U_{\lambda}(t_1,0) \nn\\
&= \sum_{r}  \Tr{}U_{\lambda}(\tau,t_1)M_r U_{\lambda}(t_1,0) U_{\lambda}^{\dg}(t_1,0) M_r^{\dg}  U_{\lambda}^{\dg}(\tau,t_1)e^{ \ln \rho_f }
   \nn\\
&= \sum_{r}  \Tr{}U_{\lambda}(\tau,t_1)M_r  M_r^{\dg}  U_{\lambda}^{\dg}(\tau,t_1)e^{ \ln \rho_f }
   \nn\\
&=  \Tr{}U_{\lambda}(\tau,t_1)  U_{\lambda}^{\dg}(\tau,t_1)e^{ \ln \rho_f }\nn\\
&=  \Tr{}e^{ \ln \rho_f }=\Tr{} \rho_f=1.
\end{align}
In the second line , we make use of $ e^{- \ln \rho_0 }\rho_0 =1$, while the cyclic property of
 trace is used in the third line. operator identity $\sum_r M_r M_r^{\dg}=1$ is used in fourth step.

We have proved that the IFT holds in the same form as in eq.(\ref{IFT}) under intermediate
 weak measurements. This can be readily generalized to the multiple intermediate weak
 measurements, which corroborates the robustness of fluctuation theorems
  under weak or generalized measurements.

\end{document}